\documentclass[conference]{IEEEtran}

\usepackage[utf8]{inputenc}
\usepackage{cite}
\usepackage{amsmath,amssymb,amsfonts}
\usepackage{graphicx}
\usepackage{textcomp}
\usepackage{xcolor}
\usepackage{url}
\usepackage{comment}
\usepackage{footnote}
\makesavenoteenv{tabular}
\makesavenoteenv{table}
\usepackage[group-minimum-digits=4,group-separator={,}]{siunitx}
\usepackage{threeparttable}
\usepackage[most]{tcolorbox}
\usepackage{pgfplots}
\usepackage{listings}
\usepackage{amsmath}
\usepackage{algorithm}
\usepackage{algpseudocode} 
\usepackage{etoolbox}
\usepackage{caption}
\usepackage{booktabs}
\usepackage{wrapfig}
\usepackage{tikz}
\usepackage{todonotes}
\usepackage{setspace}
\usepackage{enumitem}
\usepackage{hyperref}

\usetikzlibrary{shapes, arrows, calc}
\input{commands.tex}
\newcommand\copyrighttext{%
  \footnotesize \textcopyright 2025 IEEE. Personal use of this material is permitted.
  Permission from IEEE must be obtained for all other uses, in any current or future
  media, including reprinting/republishing this material for advertising or promotional
  purposes, creating new collective works, for resale or redistribution to servers or
  lists, or reuse of any copyrighted component of this work in other works.
  This work has been accepted to the 2025 IEEE/ACM International Conference on Software Engineering (ICSE) for publication. The final version is available at: \url{https://doi.org/10.1109/ICSE55347.2025.00217}}
\newcommand\copyrightnotice{%
\begin{tikzpicture}[remember picture,overlay]
\node[anchor=south,yshift=10pt] at (current page.south) 
  {\fbox{\parbox{\dimexpr\textwidth-\fboxsep-\fboxrule\relax}{\copyrighttext}}};
\end{tikzpicture}%
}

\togglefalse{peerreview}

\begin{document}

\title{From Bugs to Benefits: Improving User Stories by Leveraging Crowd Knowledge with CrUISE-AC}

\iftoggle{peerreview}{
}{  
  \author{
    \IEEEauthorblockN{Stefan Schwedt}
    \IEEEauthorblockA{
        \textit{Heriot Watt University}\\
        Edinburgh, Scotland \\
        schwedt@cruise-ac.net}
    \and
    \IEEEauthorblockN{Thomas Ströder}
    \IEEEauthorblockA{
        \textit{Fachhochschule der Wirtschaft (FHDW)} \\
        Mettmann, Germany \\
        thomas.stroeder@fhdw.de
    }
  }
}

\maketitle
\copyrightnotice

\begin{abstract}    
    Costs for resolving software defects increase exponentially in late stages. 
    Incomplete or ambiguous requirements are one of the biggest sources for defects, 
    since stakeholders might not be able to communicate their needs or fail to share their domain specific knowledge. 
    Combined with insufficient developer experience, teams are prone to constructing 
    incorrect or incomplete features. 
    To prevent this, requirements engineering has to explore knowledge sources beyond stakeholder interviews.
    Publicly accessible issue trackers for systems within the same application domain hold essential information 
    on identified weaknesses, edge cases, and potential error sources, all documented by actual users.
    Our research aims at (1) identifying, and (2) leveraging such issues to improve an agile requirements artifact known as a ``user story''.
    We present CrUISE-AC (Crowd and User Informed Suggestion Engine for Acceptance Criteria) as a fully
    automated method that investigates issues and generates non-trivial additional 
    acceptance criteria for a given user story by employing NLP
    techniques and an ensemble of LLMs.    
    CrUISE-AC was evaluated by five independent experts in two distinct business domains.    
    Our findings suggest that issue trackers hold valuable 
    information pertinent to requirements engineering.
    Our evaluation shows that 80--82\% of the generated acceptance criteria
    add relevant requirements to the user stories. 
    Limitations are the dependence on accessible
    input issues and the fact that we do not check generated criteria for being conflict-free or 
    non-overlapping with criteria from other user stories. 
\end{abstract}


\section{Introduction}
In software development, defects are a significant driver for costs. Costs for resolving defects become 
exponentially high if they are detected late in the development process \cite{Westland2004}. After a 
software system has been released, the costs for fixing a defect are up to 100 times higher compared to 
a defect detected in the early requirement phase \cite{Dawson2010}.\oldthomas{Cite only one?}

In agile software development, SCRUM is used as the major process template which proposes to note 
requirements in a textual artifact called ``user story''. A user story is typically written in the 
standardized Connextra format \textit{``As a [role], I want [action], so that [benefit]''} and acts as a reminder 
to discuss the described feature between involved roles \cite{Lucassen2016b, IEEEStandardSoftware}. Intensive 
conversation rather than documentation is supposed to ensure mutual understanding and agreement of a 
requirement’s scope between stakeholders, product owner and the development team \cite{Beck2001}. 

User stories are typically accompanied by acceptance criteria (AC). These are conditions a system must meet 
in order to fulfill a user story \cite{IEEEStandardSoftware}. They can be either written in pure natural language or 
in a semi‐structured format known as Gherkin scenarios. A Gherkin scenario follows a GIVEN-WHEN-THEN 
pattern and serves as a solid basis to derive acceptance tests \cite{Wynne2017}.

However, agile requirement engineering comes with some challenges. Stakeholders might not be able to 
communicate their needs \cite{Daneva2013} or fail to share their domain specific knowledge 
\cite{Gervasi2013}. Due to lack of knowledge, experience and inadequate requirement analysis, the 
development team is likely to build wrong or incomplete features \cite{Morales2018}. In the end, these 
results are deviations from stakeholder’s (implicit) expectations which are defined as software 
defects \cite{Fredericks1998}.

All systems within a software family share a common set of features \cite{Fredericks1998}. Observing 
and analyzing similar systems helps to get a deeper understanding about requirements \cite{Rasheed2021}. 
Considering that the current software system falls under a certain software family, public issue trackers 
can be an additional source of information about possibly hidden or yet unknown requirements. 

In this paper, we address and answer the following research question:

\textbf{How can we use issue trackers of similar systems to generate additional acceptance criteria 
for a given user story and how useful are the generated criteria?}

As an answer to the research question, we present CrUISE-AC, a novel approach to enrich user stories with automatically generated 
non-obvious acceptance criteria extracted from crowd-entered knowledge listed in public issue trackers. 
Our pipeline 
integrates a variety of AI techniques, including supervised classification\cite{supervisedClassification} 
for natural language documents,\oldthomas{Changed articles here as the previous ones seemed quite fishy. Please check. --> thanks and done} 
zero-shot instruction prompting using decoder-only LLMs\cite{Wang2019ZeroShot}
and ensemble learning\cite{Polikar2012}.

We demonstrate the proposed approach using real-world industrial user stories from two different application domains.
Our main focus is on e-commerce, from which we gathered over 300
user stories across three distinct projects. 
For this domain, over \num{165000} issues were sourced from seven different public issue trackers.
Our results reveal that 82\%\oldthomas{Was 84.5\%. Not the same value as in the abstract or evaluation -- unify across the paper and 
check correctness! --> 64,7\% for all 4 annotators, 82,3\% if 3/4 annotators agree} 
of the AC produced by CrUISE-AC represent significant enhancements to user stories, 
as determined by manual evaluation.

To evaluate the applicability of our method across different domains, we performed a similar assessment within the 
domain of content management systems (CMS). 
In this context, 80\% of all generated AC were considered useful.

Overall, the results demonstrate that issue trackers can serve as a valuable knowledge repository and 
should be utilized in conjunction with stakeholder interviews during the requirements elicitation process.
Furthermore, CrUISE-AC provides a novel approach and framework to extract, transform and utilize this data.

\section{Related Work}
Agile software development starts with the definition of a high-level project scope that is broken 
down into smaller requirements artifacts during refinements\cite{mendez2019artefacts}. These 
requirements artifacts are defined with the customer and other stakeholders\cite{cohn2010scrum}. 
Thus, stakeholders collaborate extensively with the development team to establish these 
definitions and continuously validate the product being delivered. 
According to McGrath et al.\cite{mcgrath2017stakeholder}, 
stakeholders are described in various ways within the literature, identifying 
end-users as key stakeholders.
To gather requirements and other feedback from a large distributed and heterogeneous stakeholder 
group of end-users (referred to as ``crowd''\cite{henein2006information}), 
Crowd-Based Requirements Engineering (CrowdRE) promises to provide appropriate mechanisms \cite{groen2015towards}.
Numerous studies utilize user reviews to identify and extract feature requests. SAFE uses 
part-of-speech (POS) patterns to detect features in App Reviews\cite{Johann2017safe}. Using SAFE 
as baseline, Motger et al.\cite{Motger2024tfrex} trained different transformer based models on a corpus they built from 
human-created feature tags in app descriptions.
Wu et al.\cite{Wu2021features} identified key features of an app by analyzing app descriptions and user reviews.
Moreover, the extraction of requirements from various document and text types has been explored. 
Shi et al.\cite{Shi2021} examined development emails, while 
Abualhaija et al.\cite{Abualhaija2020requirements} analyzed textual requirement specifications.
In addition to features, the processes of extracting and transforming requirements artifacts have 
been examined. Tiwari et al.\cite{Tiwari2019usecase} identified use case scenarios from textual problem 
specifications. 
Furthermore, Ngaliah et al.\cite{Ngaliah2022userstory} showed that online news can 
be utilized as a data source for creating user stories.
All studies employ natural language processing (NLP) techniques \cite{chowdhary2020natural} to identify and convert text 
fragments into artifacts that facilitate requirements engineering.\oldthomas{cite general paper on NLP here?}

Issue trackers are software tools used by organizations to interact with users and stakeholders, 
facilitating communication, documentation, and collaboration on project-related issues during the entire 
software development lifecycle\cite{Montgomery2024issuetrackers}. 
Users submit various types of issues, including \textit{feature requests}, \textit{bug reports}, and \textit{questions}, 
thereby sharing their experiences and wishes regarding the software system.
Data collected from issue trackers were the subject of investigation by several researchers, who 
developed or enhanced methods for the automated classification of 
issues\cite{Kallis2019tickettagger,pandey2017automated,siddiq2022bert}. Ruan et al.\cite{Ruan2019commit} suggested an approach to 
automatically establish missing links between issues and commits. 
To the best of our knowledge, there is no existing work that employs issue trackers to 
identify, extract, and convert their data into requirements artifacts. Furthermore, there has 
been no proposed approach to associate issue data with the existing requirements of other 
projects, thus enabling their utility in different contexts.
This paper aims to address this gap by presenting a novel concept on the rationale and 
methodology for utilizing data from issue trackers.

Throughout our experiments to match issues with user stories, we investigated the usefulness of various approaches grounded in 
information retrieval (IR)\cite{hambarde2023informationretrieval}.
In evaluating the semantic similarity between two documents of natural language text, functions like cosine can be 
employed to quantify the semantic distance after converting the documents into vector representations \cite{singhal2001informationretrieval}.
Various methods exist for converting documents into vectors, including statistically driven approaches 
(e.\,g., term frequency -- inverse document frequency (TF-IDF) \cite{roelleke2008tfidf}) or models based on neural networks 
(e.\,g., universal sentence encoder (USE) \cite{cer2018use} or Sentence-BERT (SBERT) \cite{Reimers2019SentenceBERTSE}).
As these methods are designed to retrieve the most similar documents, our experiments revealed that these approaches provided information with 
limited novelty (or even none) to a given user story. Typically, the issues returned directly mirrored the content of a user story.
Furthermore, the strategy of omitting the initial $n$ results to obtain more innovative requirements was ineffective since it was challenging to 
establish a clear beginning and end for the retrieval window.

Clustering approaches \cite{rokach2005clustering} especially utilizing DBSCAN \cite{schubert2017dbscan} 
produced good results in certain cases. However, to obtain these result, we had to use a (manually crafted) domain specific thesaurus to 
identify different words with the same meaning. Moreover, the quality of the results was highly dependent on the complexity of the 
user story, thus limiting the approach's utility. 
We first tried to classify user stories to predict when such a clustering approach would yield useful results, 
but then we found the following approach, which is both simpler, more powerful, and does not need any domain specific thesaurus.

\section{Approach}\label{sec:approach}

An overview of the proposed approach is shown in Fig.\,\ref{architecture}. 
Data objects are illustrated as rectangles featuring a curved lower edge. 
Processing steps are represented as rectangles.

The upcoming sections will detail the four major steps involved in the processing pipeline:
Preprocessing, Matching, Generation, and Evaluation.
We consider the following three inputs:

\begin{itemize}
    \item issues $I$ harvested from various issue trackers
    \item a single user story $U_i$ requiring the generation of acceptance criteria
    \item the set of already defined acceptance criteria $A^{U_i}$ for the user story $U_i$ (this set might be empty)
\end{itemize}


\newcommand{\backsep}{-2.4}
\newcommand{\backwidth}{4}
\newcommand{\duplicatedistance}{0.18}
\newcommand{\stepdistance}{1}
\newcommand{\verticaldistance}{1.8}
\newcommand{\tapewidth}{3cm}
\newcommand{\emptytape}{%
\begin{minipage}{\tapewidth}
\phantom{x}\\
\phantom{x}\\
\phantom{x}\\
\end{minipage}
}
\newcommand{\bugreportcontent}{%
\begin{minipage}{\tapewidth}
\textbf{bug report \#1}\\
- description\\
- steps to reproduce\\
\end{minipage}
}
\newcommand{\descriptioncontent}{%
\begin{minipage}{\tapewidth}
description\\
expected result\\
\phantom{x}\\
\end{minipage}
}
\newcommand{\accontent}{%
\begin{minipage}{\tapewidth}
\textbf{acceptance criterion}\\
GIVEN\\
WHEN\\
THEN
\end{minipage}
}
\newcommand{\issuelabel}{issues $I$}
\newcommand{\prepissuelabel}{\begin{minipage}{2.5cm}\begin{center}preprocessed issues $I'$\end{center}\end{minipage}}
\newcommand{\matchedissuelabel}{matched issues $I^{U_i}$}
\newcommand{\genaclabel}{\begin{minipage}{2.6cm}\begin{center}generated acceptance\\[0.6ex]criteria $A^{U_i}_g$\end{center}\end{minipage}}
\newcommand{\evalaclabel}{\begin{minipage}{2.6cm}\begin{center}relevant acceptance\\[0.6ex]criteria $A^{U_i}_r$\end{center}\end{minipage}}
\newcommand{\userstorylabel}{user story $U_i$}
\newcommand{\aclabel}{\begin{minipage}{2.7cm}\begin{center}existing acceptance\\[0.6ex]criteria $A^{U_i}$\end{center}\end{minipage}}

\begin{figure*}[bt]
    \centering
    \scriptsize
    \begin{tcolorbox}[colbacktitle=black!10!white,
                      coltitle=black,
                      colframe=black!30!white,
                      colback=white,
                      sharp corners,
                      left=1mm,
                      right=1mm,
                      top=1mm,
                      bottom=1mm]
    \resizebox{\columnwidth}{!}{%
    \begin{tikzpicture}

    \node[tape, draw, tape bend top=none, fill=white] (bug1) {\emptytape};
    \node[tape, draw, tape bend top=none, fill=white] (bug2) [below right=\duplicatedistance of bug1.north west, anchor=north west] {\emptytape};
    \node[tape, draw, tape bend top=none, fill=white, label=270:\issuelabel] (bug3)
      [below right=\duplicatedistance of bug2.north west, anchor=north west] {\bugreportcontent};
    
    \filldraw[dotted,draw=black,fill=black!10] ($(bug1.north west)+(-0.2,\backsep)$) rectangle +(\backwidth,-3);
    \filldraw[dotted,draw=black,fill=black!10] ($(bug1.north west)+(4,\backsep)$) rectangle +(\backwidth,-3);
    \filldraw[dotted,draw=black,fill=black!10] ($(bug1.north west)+(8.2,\backsep)$) rectangle +(\backwidth,-3);
    \filldraw[dotted,draw=black,fill=black!10] ($(bug1.north west)+(12.4,\backsep)$) rectangle +(\backwidth,-3);

    \node[tape, draw, tape bend top=none, fill=white] (description1) [below=\verticaldistance of bug1] {\emptytape};
    \node[tape, draw, tape bend top=none, fill=white] (description2)
      [below right=\duplicatedistance of description1.north west, anchor=north west] {\emptytape};
    \node[tape, draw, tape bend top=none, fill=white, label=270:\prepissuelabel] (description3) 
      [below right=\duplicatedistance of description2.north west, anchor=north west] {\descriptioncontent};
    
    \node (preprocessing) [above=0.1 of description1] {Preprocessing};
    
    \node[tape, draw, tape bend top=none, fill=white, label=270:\userstorylabel] (userstory) 
      [right=\stepdistance of bug1] {\begin{minipage}{\tapewidth}\begin{center}
      As a user, I want to receive an order confirmation\dots\\
      \phantom{x}\\
    \end{center}\end{minipage}};

    \node[tape, draw, tape bend top=none, fill=white] (description4) [right=\stepdistance of description1] {\emptytape};
    \node[tape, draw, tape bend top=none, fill=white] (description5) 
      [below right=\duplicatedistance of description4.north west, anchor=north west] {\emptytape};
    \node[tape, draw, tape bend top=none, fill=white, label=270:\matchedissuelabel] (description6) 
      [below right=\duplicatedistance of description5.north west, anchor=north west] {\descriptioncontent};

    \node (matching) [above=0.1 of description4] {Matching};

    \node[tape, draw, tape bend top=none, fill=white] (ac1) [right=\stepdistance of description4] {\emptytape};
    \node[tape, draw, tape bend top=none, fill=white] (ac2) [below right=\duplicatedistance of ac1.north west, anchor=north west] {\emptytape};
    \node[tape, draw, tape bend top=none, fill=white, label=270:\genaclabel] (ac3) 
      [below right=\duplicatedistance of ac2.north west, anchor=north west] {\accontent};

    \node (generation) [above=0.1 of ac1] {Generation};

    \node[tape, draw, tape bend top=none, fill=white] (ac4) [right=\stepdistance of ac1] {\emptytape};
    \node[tape, draw, tape bend top=none, fill=white] (ac5) [below right=\duplicatedistance of ac4.north west, anchor=north west] {\emptytape};
    \node[tape, draw, tape bend top=none, fill=white, label=270:\evalaclabel] (ac6) 
      [below right=\duplicatedistance of ac5.north west, anchor=north west] {\accontent};

    \node (evaluation) [above=0.1 of ac4] {Evaluation};

    \node[tape, draw, tape bend top=none, fill=white, text=white] (ac7) at (userstory -| ac4) {\begin{minipage}{\tapewidth}
      - display a ``Thank you''\\
      - the order confirmation\\\phantom{-} should be easy to print\\
      - \dots
    \end{minipage}}
      edge[dashed] (userstory);
    \node[tape, draw, tape bend top=none, fill=white, text=white] (ac8) 
      [below right=\duplicatedistance of ac7.north west, anchor=north west] {\begin{minipage}{\tapewidth}
      - display a ``Thank you''\\
      - the order confirmation\\\phantom{-} should be easy to print\\
      - \dots
    \end{minipage}};
    \node[tape, draw, tape bend top=none, fill=white, label=270:\aclabel] (ac9) 
      [below right=\duplicatedistance of ac8.north west, anchor=north west] {\begin{minipage}{\tapewidth}
      - display a ``Thank you''\\
      - the order confirmation\\\phantom{-} should be easy to print\\
      - \dots
    \end{minipage}};

    \draw (description3.east) edge[->, thick] (description4.west |- description3.east);
    \draw (description6.east) edge[->, thick] (ac1.west |- description6.east);
    \draw (ac3.east) edge[->, thick] (ac4.west |- ac3.east);

    \draw ($(bug3.south west)+(0.2,-0.12)$) edge[->, thick] ($(description1.north -| bug3.south west)+(0.2,0)$);
    \draw ($(userstory.south east)+(-0.2,-0.04)$) edge[->, thick] ($(description4.north -| userstory.south east)+(-0.2,0)$);
    \draw ($(ac9.south east)+(-0.4,-0.02)$) edge[->, thick] ($(ac4.north -| ac9.south east)+(-0.4,0)$);

    \draw[->,thick] ($(userstory.south east)+(-0.2,-0.8)$) -- ++(1.5,0) -- ($(ac1.north -| userstory.south east)+(1.3,0)$);
    \draw[->,thick] ($(userstory.south east)+(1.3,-0.8)$) -- ++(4.1,0) -- ($(ac4.north -| userstory.south east)+(5.4,0)$);

    \end{tikzpicture}
    }
    \end{tcolorbox}    
    \caption{CrUISE-AC processing pipeline}
    \label{architecture}    
\end{figure*}

\subsection{Preprocessing}\label{sec:preprocessing}

The goal of our issue preprocessing is to retain only information containing business requirements. 
For this work, we only consider issues written in English. We first eliminate pull requests (identifiable 
by their URL or description starting with a default pattern) and duplicates (i.\,e., issues with the same title 
and description). Moreover, we only keep the caption, description, and assigned labels of an issue, not making 
use of any other fields.

\begin{figure}[htb]
        \begin{tcolorbox}[colbacktitle=black!10!white,
                      coltitle=black,
                      fonttitle=\sffamily\bfseries\footnotesize,
                      colframe=black!30!white,
                      colback=white,
                      fontupper=\sffamily\footnotesize,
                      sharp corners,
                      left=1mm,
                      right=1mm,
                      flushleft title,
                      flushleft upper,
                      title=No order confirmation when changing the payment method]
        Description: \redbox{If an order is placed and e.g. Paypal is selected,}\\
        \redbox{you will receive an order confirmation}. However,\\
        \redbox{if you cancel the payment and want to complete the order}\\
        \redbox{with a new payment method in the customer account},
        no order confirmation will be sent. For this purpose, an event was created in the Flow Builder, 
        which is triggered when the payment method in the order is changed (Checkout / Order / Payment Method / Changed). 
        As an action, a mail is then sent with the mail template ``Order confirmation''. 
        The event is triggered but the mail is not generated because the ISO code is not passed.
        \\[1mm]
        Environment: Shopware 6.4.9.0
        \\[1mm]
        Steps to reproduce:
        \begin{itemize}
            \item Create a trigger in the Flow Builder with ``Checkout / Order / Payment Method / Changed'' and create an ``Order confirmation'' mail
            \item Order with e.g. Paypal until you come to the payment
            \item Cancel order
            \item Complete the order in the customer account and change the payment method beforehand
        \end{itemize}

        Expected result:
        \begin{itemize}[leftmargin=0.25cm]            
            \item[] \redbox{New order confirmation with the correct payment method}
        \end{itemize}

        Current result:
        \begin{itemize}[leftmargin=0.25cm]            
            \item[] Mail is not generated because the ISO code is missing
        \end{itemize}
    \end{tcolorbox}
    
    \caption[Sample issue]{Sample issue taken from Shopware 6}
    \label{sample_issue}
\end{figure}

Issues contain various types of information. Most of them are authored by developers 
for developers, aiming not only to convey a desired outcome (either a bug to be fixed or a new feature to be 
added) but also to provide detailed instructions on how to reproduce the mentioned issue. Many reports 
include source code to offer insights or suggestions on resolving the described issue. 
Consequently, for our approach we need to identify and eliminate information without business requirements. 
Fig.\,\ref{sample_issue} illustrates an example issue\footnote{\url{https://issues.shopware.com/issues/NEXT-20948}} 
from Shopware 6. 
Sentences enclosed in red boxes embody the actual business requirement. 

As an issue report is usually written in markdown language\cite{markdownguide}, we can easily identify and remove 
unneeded sections (such as environment, steps to reproduce, system status report) and embedded source code. 
Similarly, we eliminate URLs, image links, HTML comments and duplicate sentences.

In the last step of issue preprocessing, we identify phrases in the remaining sections which do not relate to 
any requirement. Shi et al.\cite{Shi2021} classify such phrases as trivia. Examples are 
\textit{``I hope you will find the solution''}, \textit{``Thank you''} or \textit{``I think it should be modified''}. 
We use a supervised deep learning model (see Sec.\,\ref{sec:implementation}) to identify and remove trivia phrases. 
Issues processed and reduced as described constitute the corpus $I'$.
\subsection{Matching}\label{sec:matching}
During the next phase, we associate relevant issues with the user story. 
An issue is deemed to correspond to the user story if it includes information influencing the implementation of that user story.
Usually, this means the issue contains useful 
additional requirements or outlines a bug describing an overlooked edge case. The task of matching can be reduced to a binary 
classification problem, where each user story/issue pair is assessed. The result is ``yes'' if there is a correspondence between 
the user story and the issue and ``no'' otherwise.

To match issues to user stories, we utilized decoder-only large language models (LLMs). 
These models have proven their superiority across a wide range of NLP tasks\cite{raffel2020transferlearning}
including the classification of short texts\cite{obinwanne2023GPTsentiment}.

Given the vast diversity of user stories and issues as well as the fact that our data is not labeled beforehand, supervised 
classification is impractical in this case. 
Likewise, few-shot learning turned out to be unfeasible in our experiments since we did not find good ways to generalize 
examples of such pairs.
However, decoder-only LLMs have proven their ability to generalize well across different application domains 
along with impressive zero-shot performance \cite{brown2020language}.

Thus, we pass user story/issue pairs along with a brief description of the application domain
to $k$ decoder-only LLMs using a zero-shot instruction prompt. The prompt's instructions are as follows: 
{
\sffamily
\small
Below you will find the description of an issue and a user story.
Assess, if the issue affects the functionality and role covered in the user story.
Return "yes" or "no", nothing else.
}
The full prompt is available online.\footnote{\iftoggle{peerreview}{Supplementary material: prompt\_match.txt}{\url{https://zenodo.org/records/14709846/files/prompt_match.txt}}}

Each user story $U_i$ together with the $m$-th LLM define a characteristic function 
$\mathit{match}^{U_i}_m : I' \to \{0,1\}$ with $\mathit{match}^{U_i}_m(I_j) = 1$ iff 
the $m$-th LLM returns ``yes'' to the prompt above.


An ensemble or combination of classifiers \cite{Mohandes2018ClassifierCombination} is a widely 
utilized technique in classification learning, 
which involves constructing a new classifier by integrating a set of base classifiers.
Numerous studies have demonstrated that this approach significantly enhances the classification performance compared to individual 
classifiers (see, e.\,g., \cite{bi2008combination}).
A commonality among these studies is the utilization and combination of traditional approaches. 
To the best of our knowledge, there is currently no research that explores ensemble methods 
combining the outputs of multiple decoder-only LLMs.

Our approach uses a non-adaptive ensemble algorithm \cite{delgado2022voting} using majority voting \cite{kittler1998combining}.
The process of applying and learning weights (e.\,g., as proposed in \cite{fang2024llm}) for each base model, 
along with evaluating potential improvements, is deferred to future research. 

Hence, the user story $U_i$ together with our $k$ LLMs define the overall classification result for each issue 
by the characteristic function $\mathit{match}^{U_i} : I' \to \{0,1\}$ with 
$\mathit{match}^{U_i}(I_j) = 1$ iff
$\sum\limits_{m=1}^{k} \mathit{match}^{U_i}_m(I_j) \geq \left\lceil \frac{k}{2}\right\rceil$.

Thus, the set of corresponding issues is $I^{U_i} = \{I_j \in I' \mid \mathit{match}^{U_i}(I_j) = 1\}$.

\subsection{Generation}\label{sec:generation}

Acceptance criteria for user stories are generally written either as natural language 
sentences or as semi-structured Gherkin scenarios. These criteria serve as a foundation 
for developing and supporting acceptance tests. The Gherkin format offers advantages 
over unstructured sentences by allowing for simpler automated text generation and ensuring 
traceability from acceptance criteria to executable test cases\cite{Wynne2017}. 
The Gherkin pattern consists of a brief description of the scenario itself, followed by 
a precondition \textit{GIVEN}, a trigger \textit{WHEN} and a result \textit{THEN}. 
An example of a simple Gherkin scenario is:
\begin{flushleft}
    \textbf{Scenario}: Simple search
    \begin{itemize}
        \setlength{\itemindent}{-0.5cm}
        \item[] \textbf{GIVEN} a web browser is on a search engine's page
        \item[] \textbf{WHEN} the search phrase ``cake'' is entered
        \item[] \textbf{THEN} results for ``cake'' are shown
    \end{itemize}
\end{flushleft}

To convert an issue $I_j \in I^{U_i}$ into a Gherkin-style acceptance criterion, 
we use a single LLM (this time, the output is not suitable for a simple merging mechanism like majority voting). 
The prompt contains the instruction 
{
\sffamily
\small
``Use the following issue listed under [Issue] to generate one
new acceptance criterion in Gherkin format
(GIVEN-WHEN-THEN) for the user story in [User Story].''
} 
along with the issue text $I_j$ and the user story $U_i$.
We provide the full prompt online.\footnote{\iftoggle{peerreview}{Supplementary material: prompt\_generate.txt}{\url{https://zenodo.org/records/14709846/files/prompt_generate.txt}}}

We refer to the set of generated AC for user story $U_i$ (considering all issues in $I^{U_i}$) as $A^{U_i}_g$.
An example of such an AC generation is shown in Fig.\,\ref{sample_generated_ac}.

\begin{figure}[ht]
    \begin{tcolorbox}[colback=white, boxsep=0mm, left=2mm]    
        \begin{flushleft}              
            \textbf{Input = preprocessed and matched issue $i \in I^{U_i}$}\\ 
            Session Not Expire After New Password Reset.\\
            Initially the user will login in a browser then the go with another browser 
            and changed the password for his account.
            Again he come back to his first browser but the application doesn't ask new 
            password. We need to get the new password from the user to login for different 
            browsers.\footnote{Original issue: https://github.com/opencart/opencart/issues/10449}\\
            \vspace{0.25cm}
            \textbf{Output}\\
            Scenario: Ensure session expires in other browsers after password reset                
            \begin{itemize}
                \setlength{\itemindent}{-0.5cm}
                \item[] GIVEN I am logged in on Browser A
                \item[] WHEN I change my password on Browser B
                \item[] THEN I should be prompted to log in again on Browser A with the new password
            \end{itemize}
        \end{flushleft}        
    \end{tcolorbox}    
    \caption{Gherkin scenario generated from issue}
    \label{sample_generated_ac}
\end{figure}

\subsection{Evaluation}\label{sec:assessment}

The standard for software and requirement quality \cite{IEEEStandardSoftware}
defines several characteristics for requirements to 
be \textit{complete}, \textit{consistent}, \textit{feasible}, \textit{comprehensible} and 
\textit{able to be validated}. These criteria are supported by literature and even 
extended\cite{Heck2018literaturereview,Lucassen2016aqusa}.
In this paper, we have addressed the following quality criteria so far:
\begin{itemize}
    \item \textit{completeness} by suggesting additional and potentially yet unknown acceptance criteria for a user story,
    \item \textit{feasibility} by generating acceptance criteria from issues that have been solved already in other projects,
    \item \textit{comprehensibility} and \textit{ability to be validated} by using Gherkin as a template for acceptance criteria.
\end{itemize}

\textit{Consistency} demands that requirements are unique and do not conflict with or overlap with 
other requirements in the set. We do not address the global check for conflict-free and 
non-overlapping conditions, as it necessitates considering the entire scope of all user 
stories together (we only focus on one user story at a time). However, we do address consistency 
(partially) within the scope of the one user story under consideration.

To this end, we use an LLM to determine whether the generated acceptance criteria 
introduce any new requirements beyond those already known. Additionally, 
we request that the generated acceptance criteria must contain some non-trivial and non-obvious 
knowledge. An example for a trivial requirement (which we want to eliminate) is depicted in Fig.\,\ref{fig:trivial_ac}.

\begin{figure}[htb]
    \begin{tcolorbox}[colback=white, boxsep=0mm, left=2mm]
        \begin{flushleft}
            \textbf{User story with acceptance criteria}\\
            As a logged-in user I can view and sort my favorite products.\\
            \begin{itemize}                    
                \item Clicking on ``Title A-Z'' sorts the titles alphabetically from A to Z \\
                \item Clicking on ``Title Z-A'' sorts the titles alphabetically from Z to A  \\
                \item \dots
            \end{itemize}
        \end{flushleft}            
        \begin{flushleft}        
            \textbf{Generated acceptance criterion}\\
            Scenario: Sort wishlist by product name
            \begin{itemize}
                \setlength{\itemindent}{-0.5cm}
                \item[] GIVEN I am a logged-in user
                \item[] WHEN I choose to sort my wishlist by product name
                \item[] THEN my wishlist should be displayed in alphabetical order based on product name
            \end{itemize}
        \end{flushleft}        
    \end{tcolorbox}    
    \caption{Example for trivial acceptance criterion}
    \label{fig:trivial_ac}
\end{figure}

We prompt the model to assess the quality of all new acceptance criteria along with an explanation for the taken 
decision. It contains the instruction, the user story $U_i$, the existing acceptance criteria $A^{U_i}$, one 
newly generated acceptance criterion $a \in A^{U_i}_g$,
and the issue $I_j \in I^{U_i}$ from which $a$ was generated.
The basic instruction is: {
\sffamily
\small
``For the following user story listed under [User Story] and the
supplied existing acceptance criteria listed under [Acceptance
Criteria] tell me, if the new acceptance criterion listed as [New
Acceptance Criterion] adds any unique, novel or surprising
insights and cannot be considered as common knowledge.
Your response LABEL must be either ``relevant'' or
``irrelevant''. ``relevant'' means, [New Acceptance Criterion]
adds valuable and not common knowledge.''
}\oldthomas{Changed ``criteria'' to ``criterion'' for the new criterion -- we probably need a corresponding change in the supplementary material.}
The full prompt is available online.\footnote{\iftoggle{peerreview}{Supplementary material: prompt\_evaluate.txt}{\url{https://zenodo.org/records/14709846/files/prompt_evaluate.txt}}}

The model's response for Fig.\,\ref{fig:trivial_ac} was \textit{irrelevant}. 
The decision was explained by the already existing requirement of alphabetical sorting. 
Thus, the newly generated acceptance criterion does not introduce any new conditions.

We refer to all generated acceptance criteria the model deems relevant for $U_i$ as $A^{U_i}_r$.

\subsection{Implementation}\label{sec:implementation}

As the issues do not carry a language field, we employed automatic language detection utilizing the Python package cld3~\cite{cld3}
to identify issues written in English.
To identify and filter trivia phrases during preprocessing, a supervised deep learning model was trained on a balanced corpus that 
contains \num{1916} phrases 
with an even distribution of 958 trivia and 958 non-trivia phrases.\footnote{\iftoggle{peerreview}{Supplementary material: 
trivia-trainingdata.csv}{\url{https://zenodo.org/records/14709846/files/trivia-trainingdata.csv}}}

Our objective in constructing the corpus was to analyze approximately \num{1000} random issues. 
To achieve this, we selected a random sample of 143 issues from each of the seven \mbox{e-commerce-related} issue trackers 
described in Section~\ref{sec:issues_ecommerce}.

The issues were preprocessed according to the method outlined in Section~\ref{sec:preprocessing}, except for the final step of filtering by trivia, 
which was not applied.
After preprocessing, 841 issues remained with a total of \num{2373} lines of text.
Each line was reviewed by expert E1 (see Section~\ref{sec:evaluation_ecommerce}) and labeled for trivia / non-trivia.

The final balanced corpus was split with 
a ratio of 9:1 for training and test data.
The best results were achieved by a method that integrates Robustly optimized BERT approach (RoBERTa) \cite{liu2019roberta} 
from the Transformer family and Long Short-Term Memory (LSTM) \cite{hochreiter1997long} from the Recurrent Neural Networks family. 
This combined approach has been proven to provide better results compared to training a Transformer model independently for 
various NLP classification tasks (see, e.\,g., \cite{Tan2022,mohawesh2024fake}). During experiments, 
we trained and compared BERT \cite{devlin2018bert}, RoBERTa \cite{liu2019roberta} and XLNet models \cite{yang2019xlnet}. 
The best score was achieved by the combined RoBERTa+LSTM model
with an accuracy and F1-score of 0.91 on the test data.
We provide the trained model online (hence, it can also be applied in other settings).\footnote{\iftoggle{peerreview}{link removed for anonymous-review}{\url{https://zenodo.org/records/12749484}}}

At this stage, we have not conducted a thorough evaluation of the labeling process or the model's outcomes, 
since the model serves merely as a component in the preprocessing phase.
Instead, we refer to the results of the evaluation of the entire process as detailed in Section~\ref{sec:evaluation_and_results}.

For the selection of LLMs in the matching process, we focused on smaller open-source 
models\footnote{We also evaluated some commercial GPT-models, but they demonstrated comparatively suboptimal results.}
and assessed the most popular models available for download from the Ollama 
website~\cite{ollama2024}.
These models were (presented in ascent order of billion parameters):
\texttt{phi3.5:3b}, \texttt{gemma:7b}, \texttt{gwen2:7b}, 
\texttt{mistral:7b}, \texttt{llama3:8b}, \texttt{llama3.1:8b}, 
\texttt{mistral-nemo:12b}, \texttt{phi3:14b}, \texttt{gemma2:27b}.
The best performance in our final LLM ensemble was achieved using the following five models:
\texttt{gemma2:7b}, \texttt{mistral-nemo:12b}, 
\texttt{llama3:8b}, \texttt{llama3.1:8b}, and \texttt{gemma2:27b}.\oldthomas{Do we still have the experimental data 
for this decision? Can we add it to the supplementary material? Answer: No.}

To generate Gherkin-style acceptance criteria from issues, we use OpenAI's GPT-4 Turbo 
model~\cite{openai2024gpt4turbo}. 
To get close to deterministic results, we set the \textit{temperature} (a parameter that controls 
the randomness of the model's output) to 0.
The same model is also used to check generated AC afterwards for new and non-trivial requirements compared to already existing AC.

\section{Evaluation and Results}\label{sec:evaluation_and_results}

We evaluated our approach within two different application domains. 
In our primary domain e-commerce, we collected 307 real-world user stories from three different 
projects. Additionally, over \num{160000} issues from seven different issue trackers were mined. 
To apply our results to a second domain, we obtained 34 user stories for a CMS product created by a company based in the Netherlands.
This dataset has been used in other research before and was originally created by Lucassen et al.~\cite{lucassen2016visualizing}. 
We gathered data from two distinct CMS issue trackers, mining over \num{145000} issues.

\subsection{Primary Domain: E-commerce}
We first describe the data collection followed by our evaluation methods.

\subsubsection{User stories}
For the e-commerce domain, we gained access to user stories from three distinct industry-specific closed-source projects. 
This is where our research started, since one of the authors was involved in these projects and we wanted to find improvements 
for the way how these projects were carried out. 
All projects specifically target the book industry in German-speaking countries 
and are managed by different product owners. Therefore, user stories are written by different authors.

\begin{description}[leftmargin=3mm]
    \item[Project A] defines a complete set of requirements for a B2C focused online shop 
            of a publishing house who aims to sell its own publications directly.
    \item[Project B] contains a partial set of requirements for a B2C focused online shop of a bookseller. 
    \item[Project C] includes a subset of B2C and B2B requirements for an online bookstore, 
            supplemented by an e-procurement module designed to provide information and automation for industrial customers.
\end{description}

Since AC are optional, not all user stories have some. 
Table~\ref{table_userstories} presents the number of user stories (US) we received by project along with the 
individual distribution between user stories with and without additional acceptance criteria (AC). 
We provide the dataset online.\footnote{\iftoggle{peerreview}{Supplementary material: User stories e-commerce.xlsx}{\url{https://zenodo.org/records/14709846/files/User\%20stories\%20e-commerce.xlsx}}}

\begin{table}[htb]        
    \let\TPToverlap=\TPTrlap
    \centering

    \caption{User stories by project}
    \label{table_userstories}

    \begin{threeparttable}
        \begin{tabular}{@{}p{0.485\textwidth}@{}}
            \centering        
            \begin{tabular}{c|c|c|c}
                \hline
                \bfseries Project & \bfseries \# US & \bfseries with AC & \bfseries Ratio \\
                \hline
                Project A   & 127 & \phantom{0}77 & \phantom{0}61\% \\
                Project B	& 140 & 111 & 79\% \\
                Project C	& \phantom{0}40 & \phantom{0}13 & \phantom{0}33\%\\
                \hline
                Total       & 307 & 201 & \phantom{0}65\%\\
                \hline
            \end{tabular}   
        \end{tabular}   
    \end{threeparttable}            
\end{table}

250 out of 307 user stories along with their AC were written in German language.\oldthomas{The table sums up to 279 instead of 307. We need to check the actual numbers.} 
To match them against the collected issues written in English, user stories in German had to be translated. 
For this purpose, we used the API provided by DeepL~\cite{deepl2024}.

All user stories and AC were anonymized. Company names were 
replaced by neutral terms (e.\,g., ``merchant'' or ``shop owner'') and contained URLs removed.

\subsubsection{Issues}\label{sec:issues_ecommerce}

To collect a sufficiently large dataset of issues, publicly accessible issue trackers were crawled. 
We employed data supplied by the platform BuiltWith\cite{builtwith}
to identify relevant e-commerce systems with a certain market penetration and analyzed\oldthomas{AE or BE?} this list in a top-down manner. 
We looked for systems with publicly accessible issue trackers containing a certain amount of issues. 
As of August 3rd, 2023,\oldthomas{AE or BE?} when we started our research and accessed BuiltWith, the ranking of the projects utilized\oldthomas{AE or BE?} was 
WooCommerce~(1)\footnote{\url{https://woocommerce.com}}, Magento~(3)\footnote{\url{https://magento-opensource.com}}, 
PrestaShop~(7)\footnote{\url{https://prestashop.com}}, OpenCart~(8)\footnote{\url{https://www.opencart.com}}, and 
nopCommerce~(33)\footnote{\url{https://www.nopcommerce.com}} out of 109 total systems listed.
A crawler was utilized\oldthomas{AE or BE?} to download and copy all issues that met the following criteria to a local database:

\begin{itemize}
    \item The issue's state is set to ``closed'' to capture only issues that have been reviewed by the developers.
    \item The issue is not labeled with ``cannot reproduce'', ``duplicate'', ``needs update'', ``invalid'', ``refactoring'', or ``test''.
\end{itemize}

Given that all the issue trackers of interest offered an API and supported the application of filters, the process was straightforward. 
As described in Section~\ref{sec:preprocessing}, we only downloaded the caption, description, and assigned labels of each issue, not making use of any attached 
discussion or other fields.
With this approach, over 165,000 issues could be collected that were created between June 2011 and July 2024.
For this work, we only considered artifacts written in the English language. 
As the issues do not carry a language field, we employed automatic language detection utilizing\oldthomas{AE or BE?} the Python package 
cld3\cite{cld3} which identified 120,168 issues as being written in English.
Moreover, we eliminated issues based on the following criteria:

\begin{itemize}
    \item The issue is a pull request, not a bug report or feature request.\\
          This type was identified by URLs beginning with \textit{/pull/} or descriptions starting with 
          \textit{``This issue is automatically created based on existing pull request''}.
    \item The issue was detected as a duplicate.\\ 
          Our database contained more than one issue with exactly the same title and description. 
          In this case, only the issue with the earliest creation date was kept.
\end{itemize}

Table~\ref{table_issues} presents the projects whose issue trackers were crawled along with the total 
number of collected and the final count of issues remaining after preprocessing.
We provide the full dataset of issues including both the original and preprocessed 
text online.\footnote{\iftoggle{peerreview}{Supplementary material: Issues e-commerce.xlsx}{\url{https://zenodo.org/records/14709846/files/Issues\%20e-commerce.xlsx}}}

\begin{table}[htb]
    \let\TPToverlap=\TPTrlap
    \centering    

    \caption{Data collected from issue trackers (e-commerce)}
    \label{table_issues}

    \begin{threeparttable}        
        \begin{tabular}{c|c|c}
            \hline
            \bfseries Project & \bfseries Issued downloaded & \bfseries Issues remaining\\
            \hline
            magento2\tnote{1}	    & \phantom{0}\num{33806} & \num{13081} \\
            nopCommerce\tnote{2}	& \phantom{00}\num{7131} & \phantom{0}\num{3745} \\
            OpenCart\tnote{3}	    & \phantom{0}\num{13805} & \phantom{0}\num{3690} \\
            PrestaShop\tnote{4}	    & \phantom{0}\num{31200} & \phantom{0}\num{6510} \\
            Shopware5\tnote{5}	    & \phantom{0}\num{20283} & \phantom{0}\num{2740} \\
            Shopware6\tnote{6}	    & \phantom{0}\num{15491} & \phantom{0}\num{7341} \\
            WooCommerce\tnote{7}	& \phantom{0}\num{43840} & \num{17289} \\
            \hline
            Total                   & \num{165556}& \num{54396} \\   
            \hline
        \end{tabular}
        \begin{tablenotes}\footnotesize 
            \item[1] \url{https://github.com/magento/magento2/issues}
            \item[2] \url{https://github.com/nopSolutions/nopCommerce/issues}
            \item[3] \url{https://github.com/opencart/opencart/issues}
            \item[4] \url{https://github.com/PrestaShop/PrestaShop/issues}
            \item[5] \url{https://issues.shopware.com/?products=SW-5}
            \item[6] \url{https://issues.shopware.com/?products=SW-6}
            \item[7]\url{https://github.com/woocommerce/woocommerce/issues}
        \end{tablenotes}
    \end{threeparttable}    
\end{table}\oldthomas{Unify caption style -- only first letter capitalized or title case?}
\subsubsection{Evaluation Procedure}\label{sec:evaluation_ecommerce}
Since we could not compare our results to existing approaches, we had to include a manual evaluation step 
at the end to determine the quality of our results. Hence, we had to limit the number of user stories and issues 
used in the evaluation to a feasible amount. We evaluated our approach on a random sample of 30 out of 307 user 
stories and \num{3500} out of \num{54396} preprocessed issues.
For this sample, a total of \num{1207} matching user story/issue-pairs were determined. 
For all pairs, AC were automatically generated (see Sec.\,\ref{sec:generation}) and evaluated (see Sec.\,\ref{sec:assessment}).
Fig.\,\ref{fig:ac_per_us} presents the resulting number of AC per user story.

\begin{figure}[h]
    \centering
    \begin{tikzpicture}
        \begin{axis}[
            ybar,
            symbolic x coords={1,2,3,4,5,6},
            xtick=data,
            xticklabel style={align=center, font=\footnotesize},
            xticklabels={
                $0$, $\leq10$, $\leq20$, $\leq30$, $\leq50$, $>50$
            },
            ylabel={Number of User Stories},
            xlabel={Number of AC},
            ymin=0,
            ymax=15,
            bar width=15pt,
            nodes near coords,
            height=5cm,
            width=0.46\textwidth
        ]
        \addplot coordinates {(1,4) (2,10) (3,3) (4,4) (5,4) (6,5)};
        \end{axis}
    \end{tikzpicture}    
    \caption{Number of generated AC by user story (e-commerce)}
    \label{fig:ac_per_us}    
\end{figure}

The average number of newly generated AC was 35.
The highest number of AC generated for an individual user story reached 273. 
We assume that this elevated figure is attributed to the fact that the associated user story includes specifications related to the accurate presentation of prices, including international tax rules, which is a very complex field.

We took a random sample of 10 AC for every user story with more than 10 generated AC and all generated AC for stories 
with at most 10 generated AC 
for manual inspection by four independent experts. 
In total, 198 AC for 26 user stories were reviewed.

Experts 1 (E1) and 2 (E2) are seasoned professionals in the e-commerce sector, with over 20 years of project experience and 
8 years in agile project management. 
Expert 3 (E3) is a certified Professional Scrum Master with additional certification in Professional Agile Leadership 
and seven years of experience. Expert 4 (E4) has 19 years of working experience in the software industry 
as developer, enterprise architect, agile coach, people manager, and principal consultant 
(with multiple projects both from the e-commerce and CMS domain).\oldstefan{To be filled by Thomas. TS: Done.}

Every expert reviewed the results independently and determined whether an acceptance criterion constituted 
a valid addition to the business scope of the user story, described edge cases, or covered other non-obvious scenarios.
AC that met the aforementioned conditions were tagged as ``approved''.
Table~\ref{table:manual_approval} presents the results. 

\begin{table}[ht]
    \let\TPToverlap=\TPTrlap
    \centering

    \caption{AC approval by expert (e-commerce)}
    \label{table:manual_approval}    

    \begin{threeparttable}
        \begin{tabular}{r|c|c|c|c}
            \hline
            & \bfseries E1 & \bfseries E2 & \bfseries E3 & \bfseries E4 \\
            \hline
            Approved AC   & 165 & 161 & 190 & 154 \\
            Declined AC   & \phantom{0}33  & \phantom{0}37  & \phantom{00}8   & \phantom{0}44 \\
            \hline
            Approval Rate & 83\% & 81\% & 96\% & 78\% \\
            \hline
        \end{tabular}
    \end{threeparttable}    
\end{table}

We consider an AC as accepted by the experts if it receives positive assessments from at least three out of four evaluators.
According to this definition, 82\% of the generated ACs were approved by the experts.
We offer the annotations as a downloadable 
dataset.\footnote{\iftoggle{peerreview}{Supplementary material: Evaluation e-commerce.xlsx}{\url{https://zenodo.org/records/14709846/files/Evaluation\%20e-commerce.xlsx}}}

To ensure the reliability of our annotations, we calculated the agreement rate between all annotators as 
outlined by Table~\ref{table:agreement_rate}. 

\begin{table}[ht]  
    \let\TPToverlap=\TPTrlap
    \centering

    \caption{Agreement rate of experts (e-commerce)}
    \label{table:agreement_rate}

    \begin{threeparttable}
        \begin{tabular}{c|c|c|c|c}
            \hline
            & \bfseries E1 & \bfseries E2 & \bfseries E3 & \bfseries E4 \\
            \hline
            E1      &           100.00\% & \phantom{0}92.93\% & \phantom{0}82.32\% & \phantom{0}77.27\% \\
            E2      & \phantom{0}92.93\% &           100.00\% & \phantom{0}81.31\% & \phantom{0}75.25\% \\
            E3      & \phantom{0}82.32\% & \phantom{0}81.31\% &           100.00\% & \phantom{0}77.78\% \\
            E4      & \phantom{0}77.27\% & \phantom{0}75.25\% & \phantom{0}77.78\% &           100.00\% \\
            \hline
            Average & \phantom{0}88.13\% & \phantom{0}87.37\% & \phantom{0}85.35\% & \phantom{0}82.58\% \\
            \hline
        \end{tabular}
    \end{threeparttable}
\end{table}

On average, there was an 85.85\% consensus among all experts.
We calculated Cohen's Kappa\cite{cohen1960coefficient}, 
a commonly utilized metric for evaluating inter-rater reliability\cite{viera2005understanding}.
Despite the high agreement between the raters, Cohen’s Kappa returned a relatively low value of 0.44, 
suggesting moderate inter-rater reliability\cite{landis1977measurement}. The phenomenon of high agreement, but low Kappa value
is known as Kappa paradox \cite{feinstein1990high}. Consequently, Gwet's AC1\cite{gwet2014handbook} was computed as 
another paradox-resistant indicator to provide a valid assessment of the inter-rater reliability.
Gwet's AC1 yielded a value of 0.74, indicating substantial agreement\cite{landis1977measurement}.

A complete example of a user story, its existing acceptance
criteria along with a newly generated and manually approved
criterion is displayed in Fig.\,\ref{story_ecommerce}. 

\begin{figure}[htb]
    \begin{tcolorbox}[colback=white, boxsep=0mm, left=2mm]
        \begin{small}
            \begin{flushleft}
                \textbf{User story}\\
                As a customer, I would like to register as a private user in the webshop.\\
            \end{flushleft}
            \begin{flushleft}
            \textbf{Original acceptance criteria}
            \begin{itemize}[leftmargin=0.3cm]
                    \item[--] The user must enter at least the mandatory fields for successful registration.
                    \item[--] The user will be notified if he forgets to enter mandatory fields.
                    \item[--] The user must confirm the e-mail address.
                    \item[--] The user is logged in after confirmation.
                    \item[--] The user can log in and out with his data as soon as the e-mail has been confirmed.
                \end{itemize}
            \end{flushleft}
            \begin{flushleft}
                \textbf{Newly generated acceptance criterion}\\
                Scenario: Register as a private user during checkout and proceed to confirmation page
                \begin{itemize}
                    \setlength{\itemindent}{-0.5cm}
                    \item[] GIVEN I am a new customer and I am at the checkout page
                    \item[] WHEN I complete the registration process as a private user
                    \item[] THEN I should be redirected to the checkout confirmation page instead of the account site 
                \end{itemize}
            \end{flushleft}
        \end{small}
    \end{tcolorbox}
    \caption{User story enriched with new relevant acceptance criterion (e-commerce)}
    \label{story_ecommerce}
\end{figure}

In addition, Fig.\,\ref{story_ecommerce_negative} presents a counterexample of a generated acceptance criterion 
that received approval from the automated assessment but was rejected by all the experts.

\begin{figure}[htb]
    \begin{tcolorbox}[colback=white, boxsep=0mm, left=2mm]
        \begin{small}
            \begin{flushleft}
                \textbf{User story}\\
                As a punchout-user, I want to order bundle-titles in basket, split for VAT differences all over the system.\\                
            \end{flushleft}
            \begin{flushleft}
            \textbf{Original acceptance criteria}
            \begin{itemize}[leftmargin=0.3cm]
                    \item[--] Bundles can only be ordered together.
                    \item[--] It should not be possible to delete only one part of bundles.
                    \item[--] It should not be possible to change the amount for only one part of bundles.
                    \item[--] If the print part is deleted, the online part is also deleted.
                    \item[--] Changing the amount of print part also changes the amount of the online part.
                \end{itemize}
            \end{flushleft}
            \begin{flushleft}
                \textbf{Newly generated acceptance criterion}\\
                Scenario: Ensure correct total price calculation with tax for multiple units of a product
                \begin{itemize}
                    \setlength{\itemindent}{-0.5cm}
                    \item[] GIVEN the tax rate is set to 21\%
                    \item[] WHEN the system calculates the total price including tax
                    \item[] THEN the total price should be exactly \$222.00
                \end{itemize}
            \end{flushleft}
        \end{small}
    \end{tcolorbox}
    \caption{User story with new acceptance criterion, rejected by all experts (e-commerce)}
    \label{story_ecommerce_negative}
\end{figure}

\subsection{Secondary Domain: CMS}
To evaluate the generalizability of our method in a later stage of our research, we applied it to the second domain of CMS.

We used all 34 user stories from a dataset that was originally created by Lucassen et al.\ \cite{lucassen2016visualizing}.
All user stories came without additional AC.
We provide the user stories for download since the original link is not available anymore.\footnote{\iftoggle{peerreview}{Supplementary material: User stories CMS.xlsx}{\url{https://zenodo.org/records/14709846/files/User\%20stories\%20CMS.xlsx}}}
\oldthomas{Is this set available from Lucassen et al.? If so, we should rather point to the already existing set. If there are differences (e.g., by preprocessing), we should mention this here explicitly.}

To identify relevant issue trackers, we visited the CMS usage statistics of BuiltWith\cite{builtwithcms} and conducted a search for 
publicly accessible issue trackers that offered a certain number of issues along with an API for downloading. 
To capture a broad spectrum of available CMS, we chose to examine issues from one system located in the top tier and another situated in the lower tier.
As of August, 14th 2024,\oldthomas{Why do we have two different dates? We should use the same date for both aspects.} 
BuiltWith ranked Moodle\footnote{\url{https://moodle.org}} on position 10 and Umbraco\footnote{\url{https://umbraco.org}} on position 206 
out of 287 total systems listed.\oldthomas{I keep changing BuildWith to BuiltWith -- am I wrong?}

To download and preprocess the issues for both CMSs, we reused the same mechanisms as outlined in Section~\ref{sec:issues_ecommerce}.

Table~\ref{tab:issues_cms} displays the total number of issues collected and the final count of issues remaining 
after preprocessing for both projects. 
We provide the complete dataset of issues featuring both their original and preprocessed 
text.\footnote{\iftoggle{peerreview}{Supplementary material: Issues CMS.xlsx}{\url{https://zenodo.org/records/14709846/files/Issues\%20CMS.xlsx}}}

\begin{table}[htb]
    \let\TPToverlap=\TPTrlap
    \centering

    \caption{Data collected from issue trackers (CMS)}        
    \label{tab:issues_cms}    

    \begin{threeparttable}
        \begin{tabular}{c|c|c}
            \hline
            \bfseries Project & \bfseries Issued downloaded & \bfseries Issues remaining\\
            \hline
            Moodle\tnote{1}	    & \num{133632} & \num{55752} \\
            Umbraco\tnote{2}	& \phantom{0}\num{15639}  & \phantom{0}\num{8748} \\            
            \hline
            Total               & \num{149271} & \num{64500} \\
            \hline
        \end{tabular}
        \begin{tablenotes}\footnotesize 
            \item[1] \url{https://github.com/magento/magento2/issues}
            \item[2] \url{https://github.com/nopSolutions/nopCommerce/issues}            
        \end{tablenotes}
    \end{threeparttable}    
\end{table}

We again selected a random sample of \num{3500} issues to maintain consistent parameters with those used for the primary domain.
The matching mechanism detailed in Section~\ref{sec:matching} resulted in 368 pairs of user stories and issues. 
Following the methodology discussed in Section~\ref{sec:generation}, AC were generated for all 
these pairs and evaluated automatically as described in Section~\ref{sec:assessment}. This way, 
163 AC across 28 of the 34 user stories were found to be beneficial. 
Fig.\,\ref{fig:cms_ac_per_us} details the resulting number of acceptance criteria per user story.

\begin{figure}[h]
    \centering
    \begin{tikzpicture}
        \begin{axis}[
            ybar,
            symbolic x coords={1,2,3,4,5},
            xtick=data,
            xticklabel style={align=center, font=\footnotesize},
            xticklabels={
                $0$, $\leq10$, $\leq20$, $\leq30$, $>30$
            },
            ylabel={Number of User Stories},
            xlabel={Number of AC},
            ymin=0,
            ymax=28,
            bar width=15pt,
            nodes near coords,
            height=5cm,
            width=0.46\textwidth
        ]
        \addplot coordinates {(1,6) (2,24) (3,2) (4,1) (5,1)};
        \end{axis}
    \end{tikzpicture}    
    \caption{Number of generated AC by user story (CMS)}
    \label{fig:cms_ac_per_us}    
\end{figure}

On average, 6 new acceptance criteria were generated. 
The maximum number of acceptance criteria produced for a single user story was 38.

For manual evaluation, a random sample consisting of up to 10 AC per user story was selected (for those user stories where there were 
more than 10 AC generated -- otherwise, all generated AC were considered), 
resulting in 113 user story/AC pairs in total.\oldthomas{Why don't we provide a similar chart as for the e-commerce part?}

The evaluation within the CMS domain was performed by three experts.
Experts E1 and E4 from the e-commerce domain also had significant experience in the CMS domain. Hence, we asked them again to evaluate 
our results for the CMS domain. 
Furthermore, an additional review was conducted by expert E5, a senior developer with 10 years of development experience across various CMS projects.

Table~\ref{table:cms_manual_approval} presents the outcome of the individual manual evaluation.

\begin{table}[ht]
    \let\TPToverlap=\TPTrlap
    \centering

    \caption{AC approval by expert (CMS)}
    \label{table:cms_manual_approval}

    \begin{threeparttable}
        \begin{tabular}{r|c|c|c|c}
            \hline
            & \bfseries E1 & \bfseries E4 & \bfseries E5 \\
            \hline
            Approved AC   & 86 & 86 & 88 \\
            Declined AC   & 27 & 27 & 25 \\
            \hline
            Approval Rate & 76.1\% & 76.1\% & 77.8\% \\
            \hline
        \end{tabular}
    \end{threeparttable}    
\end{table}

Again, we consider an acceptance criterion as accepted if it receives
positive assessments from the majority of the experts, specifically at least two out of the three evaluators.

Following this approach, 80\% of the generated ACs were approved by the experts. 
We provide the results of the experts' assessments as a downloadable
dataset.\footnote{\iftoggle{peerreview}{Supplementary material: Evaluation CMS.xlsx}{\url{https://zenodo.org/records/14709846/files/Evaluation\%20CMS.xlsx}}}

The agreement rate between all annotators is shown in Table~\ref{table:cms_agreement_rate}. 

\begin{table}[ht]  
    \let\TPToverlap=\TPTrlap
    \centering

    \caption{Agreement rate of experts (CMS)}
    \label{table:cms_agreement_rate}

    \begin{threeparttable}
        \begin{tabular}{c|c|c|c|c}
            \hline
            & \bfseries E1 & \bfseries E4 & \bfseries E5 \\
            \hline
            E1      &           100.00\%  & \phantom{0}68.14\%  & \phantom{0}76.99\% \\
            E4      & \phantom{0}68.14\%  &           100.00\%  & \phantom{0}80.53\%\\
            E5      & \phantom{0}76.99\%  & \phantom{0}80.53\%  &           100.00\% \\
            \hline
            Average & \phantom{0}81.71\%  & \phantom{0}82.89\%  & \phantom{0}85.84\% \\
            \hline
        \end{tabular}
    \end{threeparttable}    
\end{table}

On average, all experts reached a consensus of 83.5\%.
Cohen's Kappa yielded a value of 0.54 suggesting moderate inter-rater reliability. 
Gwet's AC1 calculated a value of 0.61, indicating substantial agreement.

A complete example of a user story taken from the CMS domain and
enriched with a newly generated and approved
criterion is displayed in Fig.\,\ref{story_cms}. 

\begin{figure}[htb]
    \begin{tcolorbox}[colback=white, boxsep=0mm, left=2mm]
        \begin{small}
            \begin{flushleft}
                \textbf{User story}\\
                As an editor, I want to crop images, so I can edit images easily without using photo editing tools, and therefore, I can work more productive.\\
            \end{flushleft}
            \begin{flushleft}
                \textbf{Newly generated acceptance criterion}\\
                Scenario: Ensure cropping is centered on the image when an aspect ratio is set
                \begin{itemize}
                    \setlength{\itemindent}{-0.5cm}                          
                    \item[] GIVEN I have uploaded an image in the grid image editor
                    \item[] WHEN I select a specific aspect ratio and initiate cropping
                    \item[] THEN the cropping tool should automatically center the crop area on the image instead of starting at the top-left corner (0,0)
                \end{itemize}
            \end{flushleft}
        \end{small}
    \end{tcolorbox}
    \caption{User story enriched with new relevant acceptance criterion (CMS)}
    \label{story_cms}
\end{figure}

Again, we provide in Fig.\,\ref{story_cms_negative} a counterexample of an acceptance criterion that was generated and automatically 
approved but subsequently rejected by all experts.

\begin{figure}[htb]
    \begin{tcolorbox}[colback=white, boxsep=0mm, left=2mm]
        \begin{small}
            \begin{flushleft}
                \textbf{User story}\\
                As an editor, I want to crop images, so I can edit images easily without using photo editing tools, and therefore, I can work more productive.\\
            \end{flushleft}
            \begin{flushleft}
                \textbf{Newly generated acceptance criterion}\\
                Scenario: Ensure content properties with ``Vary by culture'' set to true are accessible without exceptions 
                \begin{itemize}
                    \setlength{\itemindent}{-0.5cm}
                    \item[] GIVEN a content property is set to ``Vary by culture'' as true
                    \item[] WHEN the online channel manager accesses this property
                    \item[] THEN the system should allow access without throwing any exceptions
                \end{itemize}
            \end{flushleft}
        \end{small}
    \end{tcolorbox}
    \caption{User story with new acceptance criterion, rejected by all experts (CMS)}
    \label{story_cms_negative}
\end{figure}

\section{Discussion and Future Work}\label{future_research}

We have introduced CrUISE-AC as a novel approach to automatically utilize knowledge in 
issue trackers and enhance user stories with additional acceptance criteria (AC). This way, 
software development can be sped up significantly as the amount of missing or incomplete 
requirements is significantly reduced. 
Our experiments show that most of the automatically generated criteria (80--82\%) for such stories indeed add useful information. 

Future work will concentrate on how to increase this number even more. 
This includes usage of other LLMs having a higher number of parameters and evaluation of other prompt types.
Another potential improvement could be assigning and applying weights to LLMs during ensemble learning.

When running all user story/issue pairs through five different models, the matching performance demonstrates slow processing speeds for extensive issue corpora.
In our testing environment utilizing an NVIDIA GeForce RTX3060, evaluating 30 user stories against 3,500 issues across 
5 different models (totaling 525,000 prompt executions) required approximately 50 hours.
Therefore, we intend to assess whether a completely new model can be trained from scratch using the output of our LLM ensemble as training data, aiming for similar accuracy but showing significantly improved performance.

Since our approach is the first of its kind, we could not compare our findings to existing results directly and 
had to assess their quality manually. Of course, this process contains some limitations. 
In either application domain, different experts reviewed the results to determine whether the AC could generally serve as a 
valuable addition to the user story.
The experts did not take into account the scope of individual projects or the entirety of user stories and AC. 
Similarly, we did not evaluate the quality criterion of (interstory) conflict-freeness during the assessment.

Hence, assessing our approach through a field study and within actual ongoing projects is essential to determine the 
industrial applicability of CrUISE-AC.
With a larger group of domain experts, product owners, stakeholders, and developers, we will be able to either validate 
our findings or gain important insights for necessary adjustments. 

From the issues we collected, we utilize only the title, description, and labels. Nevertheless, other significant 
information may be present. The discussion surrounding an issue could encompass or disclose vital aspects of a business 
requirement. Similarly, we used closed issues only, but did not consider why an issue has been closed. Maybe it was 
rejected by the project team because of pointless requirements which will lead to pointless acceptance criteria
subsequently. Therefore, we plan to review our issue selection process and possibly consider issues with a 
corresponding code commit only.

To conclude, our research advances the state-of-the-art in requirements engineering by a new 
fully automated tool chain that supports requirements engineers, project managers, 
and project teams by paving the avenue to crowd knowledge in available issue trackers.

\oldthomas{add access date to all online sources or drop all access dates}

\bibliographystyle{IEEEtran}
\bibliography{main}

\end{document}